\documentclass[runningheads]{llncs}
\usepackage{graphicx}
\usepackage{wrapfig}
\usepackage{caption}
\usepackage{subcaption}
\usepackage[font=scriptsize]{caption}
\begin{document}
\title{A Python Extension to Simulate Petri nets in Process Mining \thanks{\scriptsize{Acknowledgments}  Funded by the Deutsche Forschungsgemeinschaft (DFG, German Research Foundation) under Germany's Excellence Strategy–EXC-2023 Internet of Production – 390621612. We also thank the Alexander von Humboldt (AvH) Stiftung for supporting our research.}}
%
%

\author{M. Pourbafrani\inst{1} , Sandhya Vasudevan\inst{2}, Faizan Zafar\inst{2},Yuan Xingran\inst{2}, Ravikumar Singh\inst{2} and Wil M. P. van der Aalst\inst{1} }
\authorrunning{Mahsa Pourbafrani et al.}
\institute{Chair of Process and Data Science, RWTH Aachen University, Germany \\
 \email{\{mahsa.bafrani,wvdaalst\}@pads.rwth-aachen.de} \and 
 {RWTH Aachen University}
 \email{\{sandhya.vasudevan, faizan.zafar,xingran.yuan,ravikumar.singh\}@rwth-aachen.de}
 }

\maketitle              
\begin{abstract}

The capability of process mining techniques in providing extensive knowledge and insights into business processes has been widely acknowledged. 
Process mining techniques support discovering process models as well as analyzing process performance and bottlenecks in the past executions of processes. However, process mining tends to be ``backward-looking" rather than "forward-looking" techniques like simulation. For example, process improvement also requires ''what-if" analyses. 
In this paper, we present a Python library which uses an event log to directly generate a simulated event log, with additional options for end-users to specify duration of activities and the arrival rate. Since the generated simulation model is supported by historical data (event data) and it is based on the Discrete Event Simulation (DES) technique, the generated event data is similar to the behavior of the real process. 

\end{abstract}
\section{Introduction}
Process mining tools provide unique capabilities to diagnose business processes existing within organizations (e.g., in transaction logs or audit trails) including discovering the running processes, as well as deviations and bottlenecks that occur or exist in the current state of the processes \cite{DBLP:books/sp/Aalst16}.
In all of the proposed tools for simulation in process mining, interaction with the user and user knowledge is an undeniable requirement for designing and running the simulation models.
Moreover, most of the approaches are dependent on external simulation tools. 
For instance, in \cite{DBLP:conf/bpm/CamargoDR19}, the proposed business process simulation technique is based on the BPMN model. All the simulation parameters with the BPMN model are put into a simulation tool such as BIMP for the simulation step. \cite{SPN} provides a comprehensive platform for modeling stochastic Petri nets, however, the connection to process mining is missing. the In \cite{DBLP:journals/is/RozinatMSA09}, the created simulation model is based on the CPN tool which requires users to have knowledge of discrete event simulation as well as \emph{Standard Machine Language} (SML) to define functions and capture the output as an event log \cite{ratzer2003cpn}. 
In \cite{howcloseGawin} an external tool, i.e., ADONIS for simulating the discovered model and parameters are used. 
\begin{figure}
    \centering
    \includegraphics[width=0.9\textwidth, height=0.15\textheight]{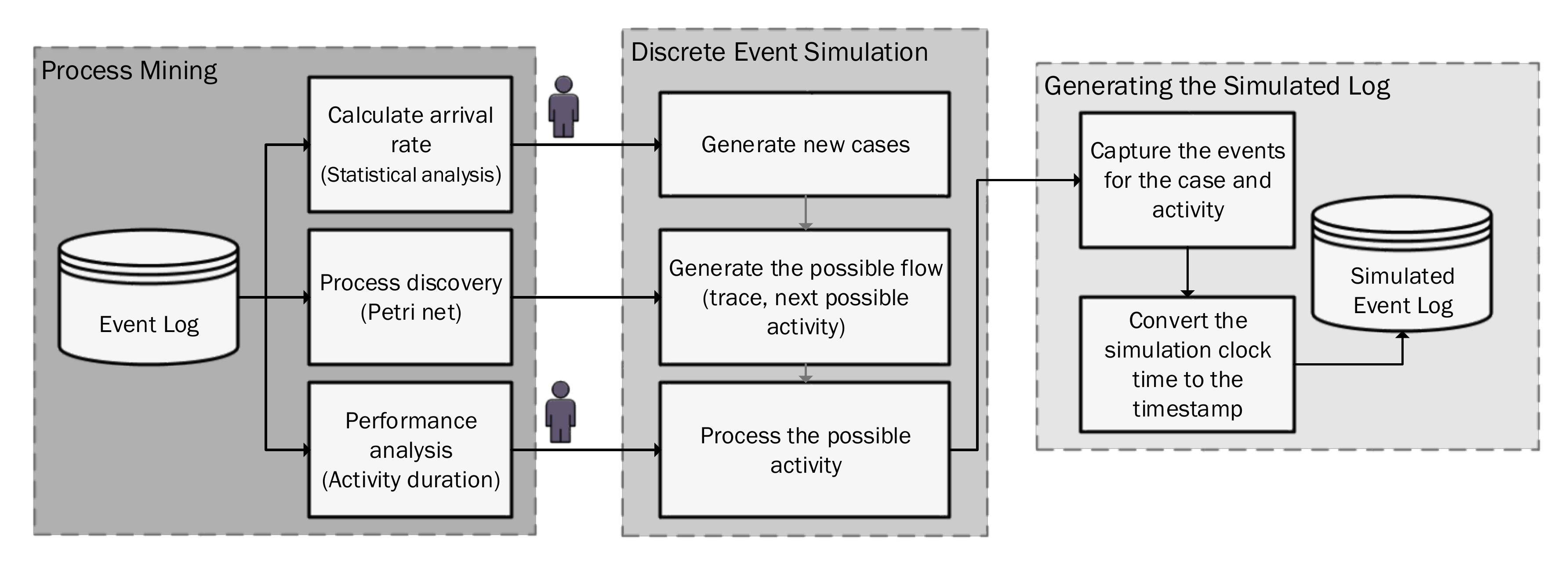}
    \caption{The general framework for discrete event simulation in process mining. The automatic generation of simulation models and the corresponding simulated event logs is possible by starting with an event log, extracting the process model and the performance information, generating random cases, and finally converting the processed activities in the form of events. The user annotations indicate the options for the user to simulate the process with user-defined parameters.}
    \label{fig:Main_Framework_PMSIM}
    \vspace{-8mm}
\end{figure}
It should be noted that oftentimes the user does not need to have in-depth knowledge of the process so as to simulate it which holds for the most of commercial tools such as \emph{Protos}, \emph{Any Logic} and \emph{ARENA}. For instance, when the user only needs to know how the process will behave if the average arrival rate increases to 5 minutes, i.e., every 5 minutes a new case arrives. 

In process mining, the above-mentioned requirements can be addressed by the concept of Discrete Event Simulation (DES) \cite{DBLP:conf/scsc/Aalst18}. DES for business processes has been developed in \emph{Java} as a plugin of \emph{ProM} \cite{VanDongen2005}. However, custom options such as the ability to change the duration of activities for future performance analyses are missing.
Approaches such as in \cite{DBLP:journals/is/Rogge-SoltiW15} uses the same idea in Java, including some drawbacks, e.g., a fixed duration for case generation step. The generated cases do not have any time overlap, which is not the case in reality. 
Work such as \cite{DBLP:conf/bpm/PufahlW17} tries to generate a business simulation model for business processes which relies on the user domain knowledge. 
\cite{MsimModelConstructionMartinDC16} describes a range of modeling tasks that should be considered when creating a realistic business process simulation model.

Existing process mining tools provide users with a visual representation of process discovery and performance analyses using event data in the form of event logs.
Therefore, an approach is needed to play out reality and generate the exact behavior which makes further analyses in process mining possible. 
Moreover, the option to extend the library as an open-source tool is easily provided. User options to add capacity to the activities and to extend the case production for different times of the day and week can be implemented.

Research work such as \cite{DBLP:conf/otm/PourbafraniZA19} and \cite{MahsaBIS} use aggregated simulation which is useful for what-if analyses in a high-level decision-making scenario \cite{DBLP:conf/ihsi/PourbafraniZA20}. The \emph{PMSD} tool represents the aggregated approach and generates a simulation model at a higher level of detail \cite{mahsaToolPMSD}. 

In this paper, we introduce an easy-to-use open-source Python-based application that connect the provided process mining environment in Python \emph{PM4Py} \cite{DBLP:pm4py} to the general simulation techniques in Python, \emph{Simpy} \footnote{https://simpy.readthedocs.io}. 
The latter library is used for discrete event simulation and handles the required system clock in DES. The automatically designed simulation model can be configured with user-defined duration for the activities and arrival rate. The final output is an event log based on the given number of cases that can be used further for process mining analyses.
The designed framework of the tool is shown in Fig. \ref{fig:Main_Framework_PMSIM}. It is designed on the basis of three main modules; process mining, simulation, and transformation of the generated events into an event log.

\section{PNSIM}

Event logs comprise events where each event refers to a \emph{case} (process instance), an \emph{activity}, a \emph{timestamp}, and any number of additional attributes (e.g., costs, resources, etc.). A set of events forms an event log which can be used in process mining analyses. As shown in Fig. \ref{fig:Main_Framework_PMSIM}, our approach starts with applying process mining techniques on the original event log. Therewith a process model is discovered in the form of a Petri net which presents possible flows of activities for the cases. Subsequently, performance analyses provide the case arrival rate including the business hours and the average duration of the activities. This information makes the automatic generation of process instances based on the past executions of processes possible. 

\begin{wrapfigure}{r}{0.495\textwidth}
    \centering
    \vspace{-10 mm}
    \includegraphics[width=0.5\textwidth,height=0.2\textheight ]{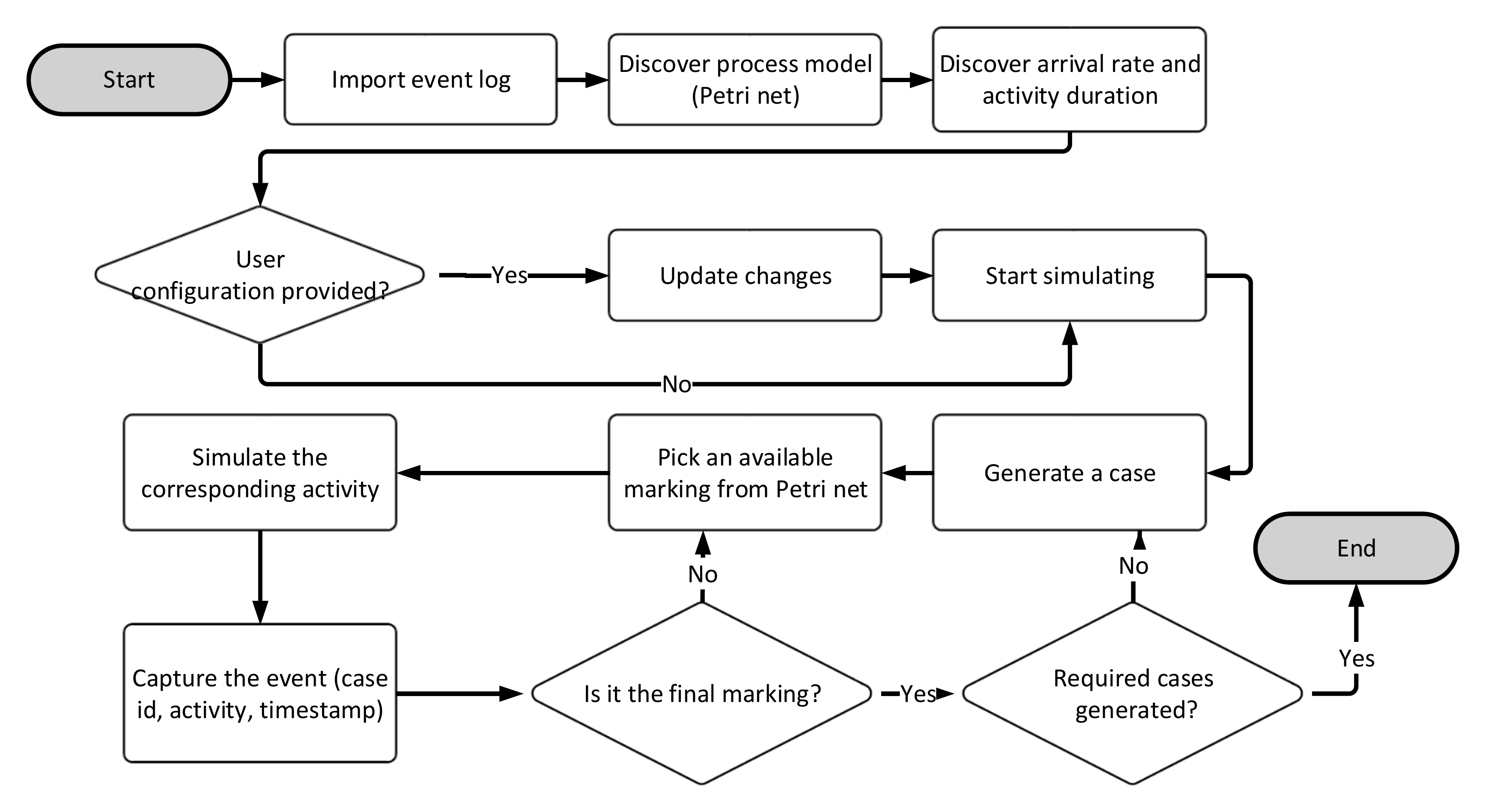}
    \caption{The flowchart of the integrated discrete event simulation of the processes using process mining. Each activity runs if it is available and the clock of the simulation gets updated for every new event. New events are a newly arrived case, the end of processing an activity for a case, or the start of the processing of an activity for a case.}
    \label{fig:flowchart}
    \vspace{-9 mm}
\end{wrapfigure}
We aim to provide a simulation model and the corresponding simulated event log as close to reality as possible. To do so, we perform the following preprocessing steps in the process mining module:
\begin{itemize}
    \item Process discovery:
    \begin{itemize}
        \item \emph{Maximum length of traces}: The presence of loops in the process models (Petri nets) makes the generation of long unrealistic traces possible. By identifying and replacing the maximum length of traces, we limit the possibility of the execution of unrealistic loops for the simulated cases. 
    \end{itemize}
    \item Performance analyses:
    \begin{itemize}
        \item \emph{Arrival rate calculation}: The business hours are considered by default in calculating the average arrival rate. Moreover, we learn the inter-arrival time distribution from the actual arrival times. 
        The detected distribution is used in the simulation step. 
        \item \emph{Activity duration}: By removing outliers from the set of duration for each activity, we provide more robust values for the duration of activities. 
    \end{itemize}

\end{itemize}
Using the distribution of activities’ duration, we implicitly consider the average duration of resources’ time without extracting the resource pool. This aggregated calculation includes the behavior of resources for handling each activity.  

Next is the simulation module in which we generate new cases. In extracting the arrival rate of cases, i.e., the duration of time for a new case to arrive, we include the business hours in the calculation of the arrival rate to obtain an accurate value. The next step is to discover how the cases are handled in the process w.r.t. the service time of each activity and the possible flow of activities that each case can take. Based on the presence of the start and complete timestamps, the value of the average duration of each activity is captured. The discovered Petri net also is used for generating a possible flow of activities. 
The provided user options to interact with and modify the simulation process are the following functions:
\begin{itemize}
    \item \emph{Activity duration} generates the random values based on the extracted values for each activity and the corresponding distribution. The user is able to change the parameters of the distribution .
    \item \emph{Arrival rate} uses a normal distribution for generating new cases and the user is able to change the average arrival rate for the simulated log.
    \item \emph{Case generator} produces random cases based on the provided number of cases by the user. It determines the terminating point of the simulation. 
\end{itemize}
The final module is designed to transform the simulated events for the generated cases into event logs. The discrete event simulation clock is converted to the real timestamp and each activity is recorded for the cases in the real timestamp. 
The flow chart of the simulation module of our tool is shown in Fig. \ref{fig:flowchart}. 
After each new generated case, it checks the condition whether the number of cases provided by the user is met. Accordingly, it follows up with processing the picked marking from the Petri net. Either the provided outputs by the process mining module or user parameters are used to start the simulation. By selecting the available activity from the Petri net, the simulation module checks whether the previous process of the activity has finished. In the last step, after performing each possible event (generating a new case or processing of an activity) the simulation clock gets updated and the data is captured. Since the simulation technique considers the capacity of each activity, the concept of queuing is implicitly covered in the simulated event log. When an activity with full capacity, i.e., processing other cases, is selected for the current case, the case is in the waiting state which is shown in the performance analyses of the event log.

\section{Tool Maturity}
The source code of our tool, a tutorial, and a screen-cast are publicly available. \footnote{https://github.com/mbafrani/AutomaticProcessSimulation}
The tool has been used in multiple academic projects to simulate a process model in different situations and generate different event logs.
\begin{figure}[bt]
\vspace{-4mm}
    \centering
    
    \begin{subfigure}{0.49\textwidth}
 \centering
    \includegraphics[width=\textwidth,height=0.25\textheight]{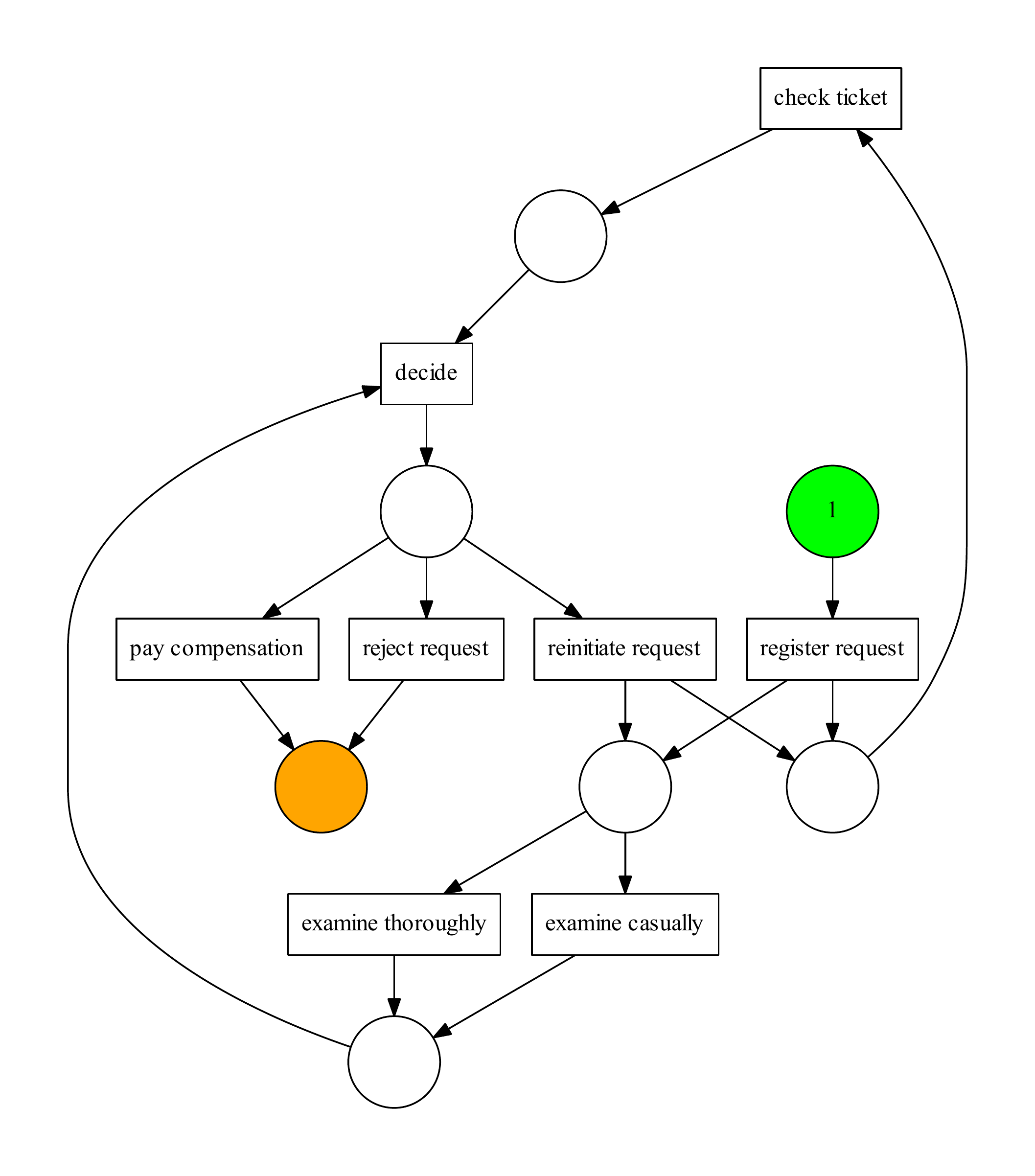}
\end{subfigure}
\begin{subfigure}{0.49\textwidth}

\centering
    \includegraphics[width=\textwidth,height=0.2\textheight]{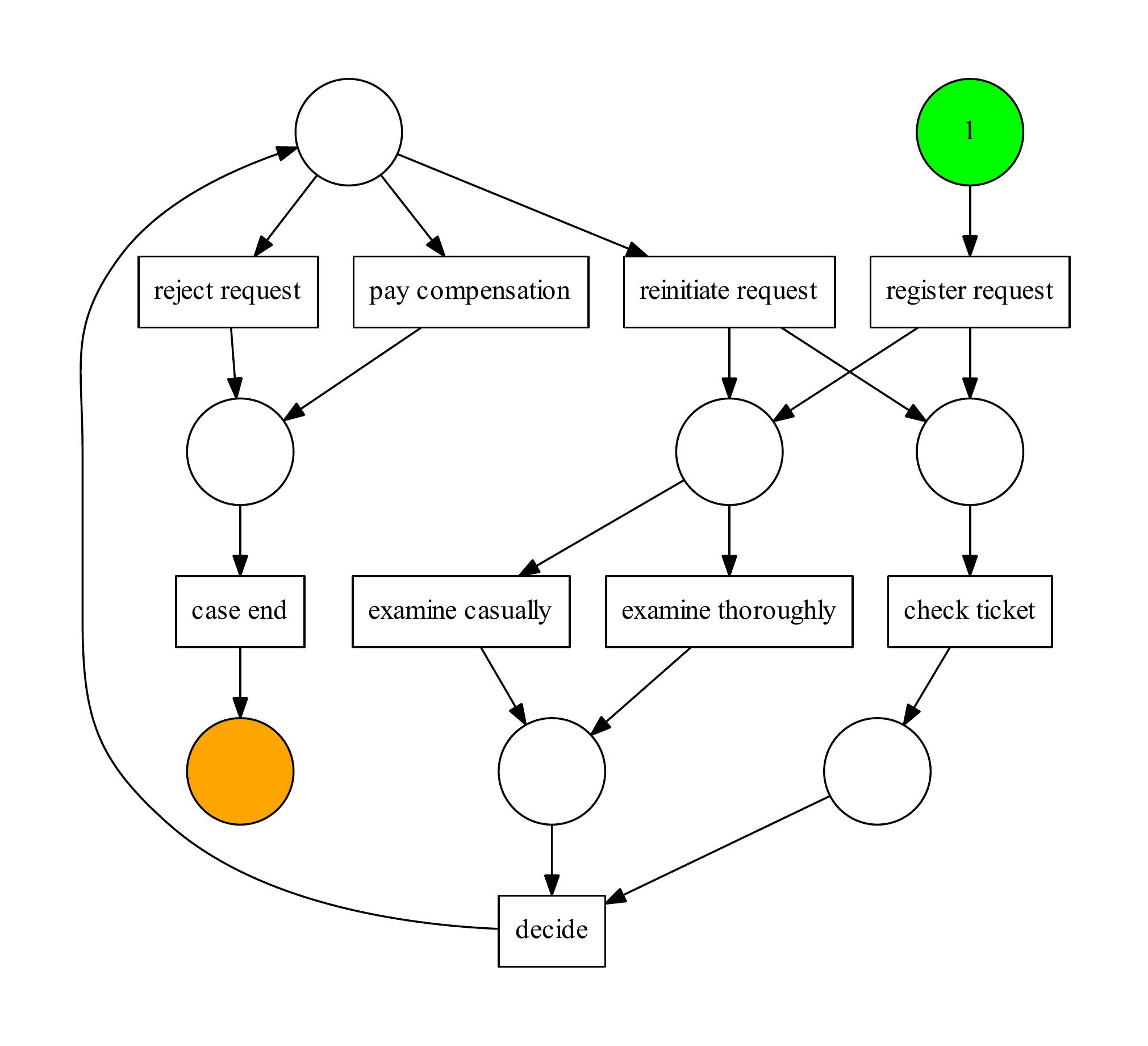}
\end{subfigure}

 \caption{The discovered process model of the example event log using Petri net notation. It includes 8 unique activities and represents the process of handling requests in an organization (a).The discovered process model of the simulated event log using Petri net notation. Our tool generates the simulated event log directly from the original event log, which captures both time and activity flow features of the original process (b).}
 \label{fig:PNsimReal}
\vspace{ -8 mm}
\end{figure}
For instance, for the purpose of time series analyses, different arrival rates for the same process have been selected and the tool event logs are generated. We use a sample case study to demonstrate the steps and usability of our tool. 
\begin{wrapfigure}{r}{0.5\textwidth}
\vspace{-2 mm}
    \centering
    \includegraphics[width=0.41\textwidth,height=0.12\textheight]{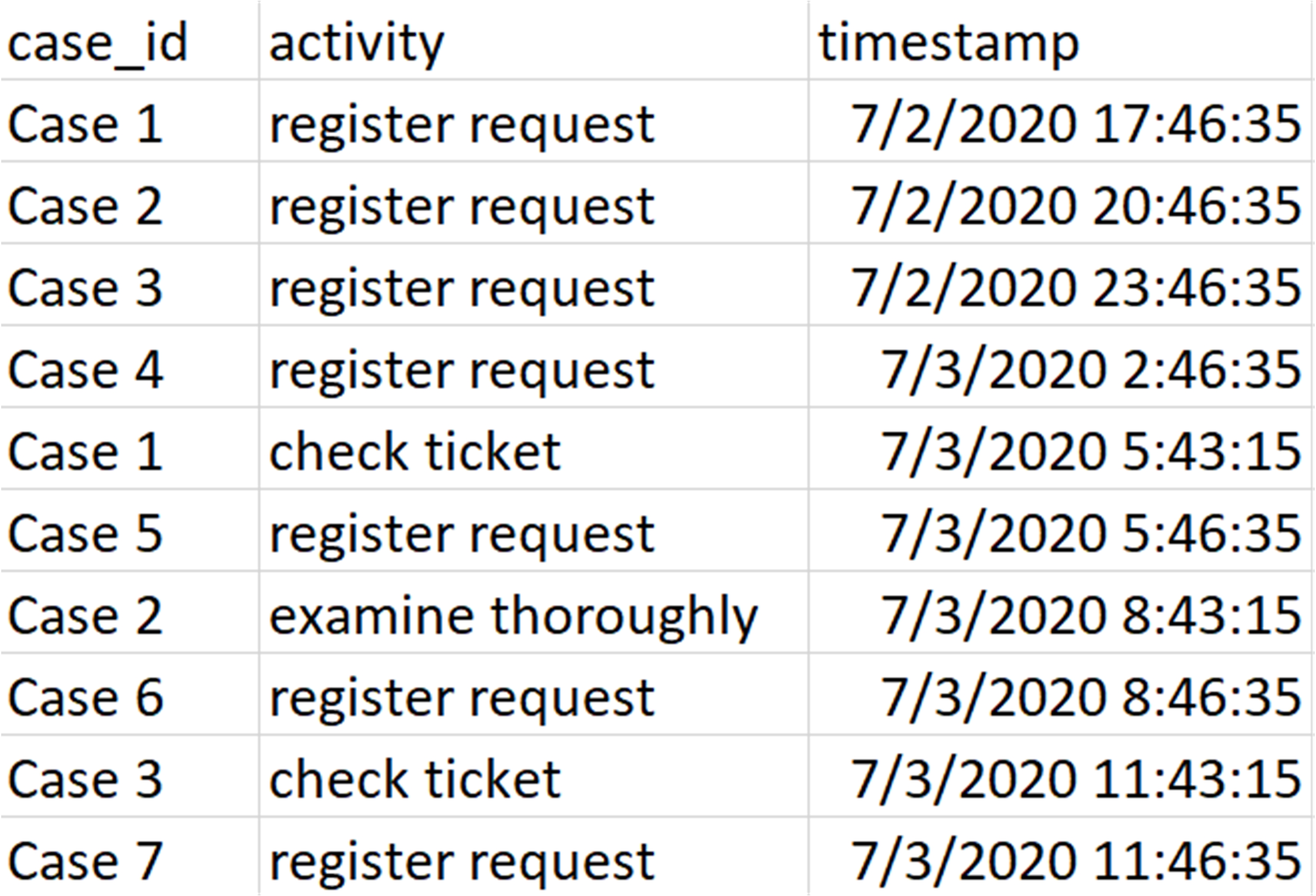}
    \caption{Part of the simulated event log for the example event log which is generated in the \emph{.csv} format. It includes the main attributes of an event log, case id, activity, and timestamp.  }
    \label{fig:simulated_event_log}
    \vspace{ -8 mm}
\end{wrapfigure}
Figure \ref{fig:PNsimReal}(a) shows a sample process model of the example event log in the form of a Petri net. We use the sample event log and simulate the process for 1000 cases. Using the same process discovery algorithm for the simulated event log result in the same model including concurrences in the model as shown in Fig. \ref{fig:PNsimReal}(b). The discovered model shows that our tool is able to mimic the process and simulate the model including the time aspects of the process. 
Part of the simulated log is shown in Fig. \ref{fig:simulated_event_log}. The simulated event log has the main attributes of an event log. It captures the \emph{case id} which is increased incrementally to the defined number by the user, \emph{activity} names, and the corresponding complete time as \emph{timestamp}. 

\section{Conclusion}
Techniques for past analyses of processes in organizations are well-supported in existing academic and commercial process mining tools. However, future analyses for business processes are not fully covered in the current tools. Commonly used options either need knowledge of simulation techniques and modeling, high interaction with users or are not accurate enough since they are not supported by real event data. In this paper, we presented the tool which directly uses the event data of a process in the form of an event log and simulates the process with the automatically extracted values as well as user-defined input. The tool is designed to simulate the processes in different scenarios. Since the simulation module is based on the discrete event simulation technique, the simulated event log includes the same behavior as the real event log.

\bibliographystyle{ieeetr}

\bibliography{Reference}
\vspace{12pt}
\end{document}